# 後量子密碼學與橢圓曲線密碼學混合憑證方案—以車聯網安全憑證管理系統為例

Abel C. H. Chen　Bon-Yeh Lin

Information & Communications Security Laboratory,
Chunghwa Telecom Laboratories
Corresponding Author: Abel C. H. Chen (chchen.scholar@gmail.com)

**摘要**

　　由於現行的車聯網安全憑證管理系統(Security Credential Management System, SCMS)標準採用的非對稱式密碼學是橢圓曲線密碼學(Elliptic-Curve Cryptography, ECC)，而橢圓曲線密碼學將可能被量子計算所破解，所以車聯網安全憑證管理系統受到量子計算攻擊的威脅。然而，目前雖然美國國家標準與技術研究院(National Institute of Standards and Technology, NIST)已經評選出後量子密碼學演算法(Post-Quantum Cryptography, PQC)標準，但現行的後量子密碼學演算法將可能存在公鑰長度較大、簽章長度較大、或是簽驗章效率較差等議題，可能無法完全符合車聯網應用需求。有鑑於車聯網存在封包長度、簽章和驗章效率、安全等級、車輛隱私等挑戰，本研究提出後量子密碼學與橢圓曲線密碼學混合憑證方案，結合後量子密碼學和橢圓曲線密碼學各自的優勢，建立克服車聯網挑戰的解決方案。運用後量子密碼學來建立符合抗量子計算的安全等級，並且通過橢圓曲線密碼學建立匿名憑證和減少封包長度以符合車聯網傳輸需求。在實驗中，本研究採用中華電信的安全憑證管理系統和公信電子的車載設備(On-Board Unit, OBU)在新北淡海新市鎮場域進行實證。比較現行的各種後量子密碼學(包含 Dilithium、Falcon、SPHINCS+)與橢圓曲線密碼學混合憑證方案的效能，並提供可實際落地的解決方案。

*關鍵詞：安全憑證管理系統、後量子密碼學、橢圓曲線密碼學、註冊憑證、授權憑證*

# Hybrid Scheme of Post-Quantum Cryptography and Elliptic-Curve Cryptography for Certificates — A Case Study of Security Credential Management System in Vehicle-to-Everything Communications

Abel C. H. Chen　Bon-Yeh Lin

Information & Communications Security Laboratory,
Chunghwa Telecom Laboratories

**Abstract**

　　Due to the current standard of Security Credential Management System (SCMS) for Vehicle-to-Everything (V2X) communications using asymmetric cryptography, specifically

Elliptic-Curve Cryptography (ECC), which may be vulnerable to quantum computing attacks. Therefore, the V2X SCMS is threatened by quantum computing attacks. However, although the National Institute of Standards and Technology (NIST) has already selected Post-Quantum Cryptography (PQC) algorithms as the standard, the current PQC algorithms may have issues such as longer public key lengths, longer signature lengths, or lower signature generation and verification efficiency, which may not fully meet the requirements of V2X communication applications. In view of the challenges in V2X communication, such as packet length, signature generation and verification efficiency, security level, and vehicle privacy, this study proposes a hybrid certificate scheme of PQC and ECC. By leveraging the strengths of both PQC and ECC, this scheme aims to overcome the challenges in V2X communication. PQC is used to establish a security level resistant to quantum computing attacks, while ECC is utilized to establish anonymous certificates and reduce packet length to meet the requirements of V2X communication. In the practical experiments, the study implemented the SCMS end entity based on the Chunghwa Telecom's SCMS and the Clientron's On-Board Unit (OBU) to conduct field tests in Danhai New Town in New Taipei City. The performance of various existing hybrid certificate schemes combining PQC (e.g., Dilithium, Falcon, and SPHINCS+) and ECC is compared, and a practical solution is provided for V2X industries.

# 1.前言

近年來,隨著無人駕駛汽車和駕駛輔助系統的發展,汽車聯網和其他車載設備(On-Board Unit, OBU)或路側設備(Road-Side Unit, RSU)通訊的需求也越加迫切。因此,北美提出專用短距離通訊(Dedicated Short-Range Communications, DSRC)(Kenney, 2011)、歐洲提出第五代通訊新廣播車聯網(5G New Radio-V2X, 5G NR-V2X)(Garcia et al., 2021)及第四代長期演進車聯網(4G Long Term Evolution-V2X, 4G LTE-V2X)(ETSI, 2020)等車聯網(Vehicle-to-Everything, V2X)通訊網路,建立車輛與其他設備間的通訊,並且陸續在許多場域進行實測。

為保障車聯網的安全通訊和保護用戶隱私,北美主要採用國際電機電子工程師學會(Institute of Electrical and Electronics Engineers, IEEE)提出的安全憑證管理系統(Security Credential Management System, SCMS)標準(ITSC of IEEE VTS, 2022),而歐洲主要採用歐洲電信標準協會(European Telecommunications Standards Institute, ETSI)提出的協同智慧運輸系統憑證管理系統(Cooperative Intelligent Transportation System (C-ITS) Credential Management System, CCMS)標準(Berlato, Centenaro & Ranise, 2022)。雖然美規和歐規的標準有些許差異,但在車聯網封包簽章和驗章採用的非對稱式密碼學皆為橢圓曲線密碼學(Elliptic-Curve Cryptography, ECC)。然而,已有研究指出 Shor 量子演算法將可能破解橢圓曲線密碼學(Shor, 1997),所以量子計算將可能對現行的車聯網安全造成威脅。

有鑑於量子計算攻擊造成的威脅,美國國家標準與技術研究院(National Institute of Standards and Technology, NIST)在 2022年已經徵求並評選出後量子密碼學(Post-Quantum Cryptography, PQC)演算法標準,包含 Kyber、Dilithium、Falcon、以及 SPHINCS+等可抗量子計算攻擊的密碼學演算法(Alagic et al., 2022)。然而,雖然這些後量子密碼學演算法可以抵抗量子計算攻擊,但可能存在公鑰長度較大、簽章長度較大、或是簽驗章效率較差等議題,可能無法完全符合車聯網應用需求。因此,本研究先整理出車聯網安全憑證管理系統主要挑戰,再根據這些挑戰設計可行的解決方案。

車聯網安全憑證管理系統主要挑戰條列如下:

(1). **封包長度**:由於 IEEE 1609.3 標準中定義車載環境無線存取短訊息(Wireless Access in the Vehicular Environment (WAVE) Short Message, WSM)的預設長度為 1400 位元組(bytes)(ITSC of IEEE VTS, 2021),所以

不論在第五代通訊新廣播車聯網(Garcia et al., 2021)、第四代長期演進車聯網(ETSI, 2020)、或專用短距離通訊(Kenney, 2011)網路傳輸時車輛傳送的封包需小於 1400 位元組，超過封包長度上限將無法被傳送。

(2). **簽章和驗章效率**：在北美車聯網概念性驗證(Proof of Concept, POC)場域中，要求車載設備每 100 毫秒發送一筆帶有基本安全訊息(Basic Safety Message, BSM)的安全協定資料單元(Secure Protocol. Data Unit, SPDU)，並且每一筆安全協定資料單元都需要加入簽章(V2X Core Technical Committee, Society of Automotive Engineers (SAE), 2020)。除此之外，車載設備需接收附近其他車載設備或路側設備發送的安全協定資料單元，並且對安全協定資料單元裡面的簽章資訊進行驗章。因此，在車聯網環境中需求有足夠快速的簽章和驗章效率，以達到車聯網封包的時效性。

(3). **安全等級(Security Level)**：美國國家標準與技術研究院早期訂定的安全強度(Security Strength)(Barker, 2020)較不適用於抵抗量子計算的安全分析，所以重新定義安全等級(Security Level)來衡量後量子密碼學演算法(NIST, 2023)。然而，安全等級越高，將可能導致公鑰和憑證長度越大、簽章和驗章計算時間越久。因此，需在符合安全等級且符合封包長度及簽驗章效率限制的平衡下，設計解決方案。

(4). **車輛隱私**：在車聯網環境中，車輛多數為個人車輛，所以存在個人移動隱私的議題。為避免曝露個人和車輛的隱私，IEEE 1609.2.1 標準設計蝴蝶金鑰擴展機制搭配橢圓曲線密碼學隱式憑證(implicit certificate)，產製包含蝴蝶公鑰的假名憑證(pseudonymous certificate)，並且避免被攻擊者從蝴蝶公鑰(butterfly public key)反推毛蟲公鑰(caterpillar public key)，達到匿名性(anonymity)和保護隱私(ITSC of IEEE VTS, 2022)。然而，美國和加拿大合作研究指出現行後量子密碼學標準演算法無法做到隱式憑證(Bindel & McCarthy, 2023)，所以如何在結合後量子密碼學的基礎上設計匿名憑證為車聯網重要的挑戰之一。

因此，本研究提出後量子密碼學與橢圓曲線密碼學混合憑證方案，結合後量子密碼學和橢圓曲線密碼學各自的優勢，建立克服車聯網挑戰的解決方案。運用後量子密碼學來建立符合抗量子計算的安全等級，並且通過橢圓曲線密碼學建立匿名憑證和減少封包長度以符合車聯網傳輸需求。

本文分為五個章節。第二節將討論現行的 IEEE 1609.2 標準(ITSC of IEEE

VTS, 2023)和 IEEE 1609.2.1 標準(ITSC of IEEE VTS, 2022)的憑證申請流程、憑證格式、以及安全協定資料單元格式。第三節以 IEEE 1609.2 和 IEEE 1609.2.1 標準為基礎，提出後量子密碼學與橢圓曲線密碼學混合憑證方案，強化現行安全標準不足處，並符合現行通訊需求。第四節將對本研究提出的方案進行實驗比較，分別從憑證長度與安全協定資料單元長度、簽章和驗章效率、加密和解密效率等面向討論。最後，第五節將總結本研究的貢獻和討論未來可行的研究方向。

## 2. 文獻探討

本研究主要參考北美標準，所以著重於 IEEE 1609.2 標準(ITSC of IEEE VTS, 2023)和 IEEE 1609.2.1 標準(ITSC of IEEE VTS, 2022)。以下將整理和討論 IEEE 1609.2 和 1609.2.1 標準中的安全憑證管理系統系統架構、憑證申請流程、憑證格式、以及安全協定資料單元格式。

### 2.1 安全憑證管理系統系統架構

IEEE 1609.2.1 標準定義的安全憑證管理系統，包含有根憑證中心(Root Certificate Authority, RCA)、中繼憑證中心(Intermediate Certificate Authority, ICA)、授權憑證中心(Authorization Certificate Authority, ACA) (IEEE 1609.2.1-2022 版名稱)/假名憑證中心(Pseudonym Certificate Authority, PCA)(IEEE 1609.2.1-2020 版名稱)、註冊憑證中心(Enrollment Certificate Authority, ECA)、登錄中心(Registration Authority, RA)、以及終端設備(End Entity EE)等，如圖 1 所示(ITSC of IEEE VTS, 2022)。

在安全憑證管理系統中，由根憑證中心自簽發根憑證中心本身的憑證，並且根憑證中心的憑證需被加入憑證信任列表(Certificate Trust List, CTL)，通過選舉人(Elector)對憑證信任列表進行簽章後，憑證信任列表和根憑證中心才能成為安全憑證管理系統的信任根源(trust anchor)。在憑證串鏈檔(Certificate Chain File, CCF)中將包含憑證信任列表、根憑證中心憑證、中繼憑證中心憑證、授權憑證中心憑證/假名憑證中心憑證、以及註冊憑證中心憑證。由根憑證中心簽發中繼憑證中心憑證，並且由中繼憑證中心簽發授權憑證中心憑證/假名憑證中心憑證、以及註冊憑證中心憑證。終端設備(包含車載設備(On-Board Unit, OBU)和路側設備(Road-Side Unit, RSU))可以取得憑證串鏈檔，並且驗證整個憑證串鏈，驗證全部簽章無誤後可以信任憑證串鏈檔中各個憑證中心簽發的憑證(ITSC of IEEE VTS, 2022)。

在管理和應用上，終端設備憑證根據用途可以分為註冊憑證(enrollment

certificate)和授權憑證(authorization certificate)/假名憑證(pseudonym certificate)。其中，由註冊憑證中心簽發終端設備的註冊憑證，並且由授權憑證中心/假名憑證中心簽發終端設備的授權憑證/假名憑證(ITSC of IEEE VTS, 2022)，具體流程將於 2.2 節中詳述。

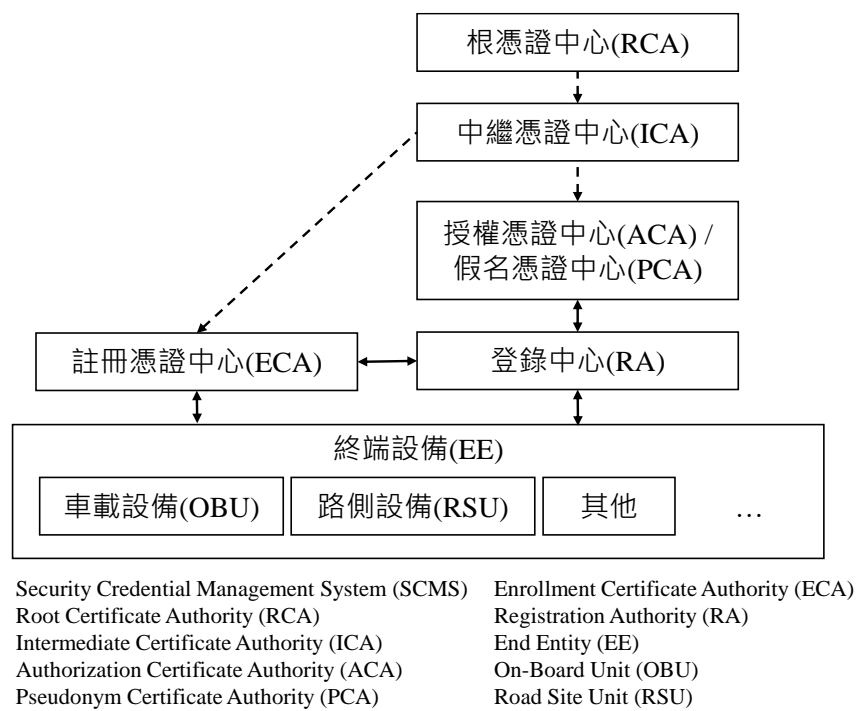

圖 **1** 安全憑證管理系統系統架構**(ITSC of IEEE VTS, 2022)**

## 2.2 憑證申請流程

IEEE 1609.2.1 標準定義的憑證申請流程包含有註冊憑證申請流程和授權憑證/假名憑證申請流程，如圖 2 所示(ITSC of IEEE VTS, 2022)。

在註冊憑證申請流程前，終端設備在生產出廠時需先預存權威識別碼(canonical ID)和權威私鑰(canonical private key)，並且註冊憑證中心需儲存終端設備的權威識別碼(canonical ID)和權威公鑰(canonical public key)。當終端設備啟用時，可以執行註冊憑證申請流程，由終端設備隨機產生一組註冊金鑰對(enrollment key pair)，並把權威識別碼和註冊公鑰放到 EeEcaCertRequest，以及用權威私鑰對 EeEcaCertRequest 內容進行簽章，再發送 EeEcaCertRequest 給註冊憑證中心。註冊憑證中心收到 EeEcaCertRequest 後，用權威識別碼查詢對應的權威公鑰，以權威公鑰對 EeEcaCertRequest 驗章；當驗章成功後，代表是合法的終端設備，並且將終端設備的註冊公鑰放到註冊憑證，並且簽發註冊憑證和把註冊憑證放到 EcaEeCertResponse 回傳給終端設備(ITSC of IEEE VTS, 2022)。

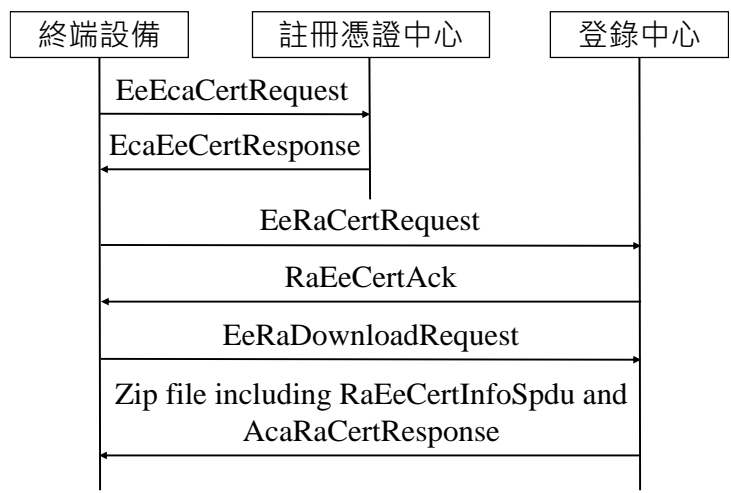

**圖 2　憑證申請流程(ITSC of IEEE VTS, 2022)**

當終端設備收到 EcaEeCertResponse 後，可以執行授權憑證/假名憑證申請流程，由終端設備根據蝴蝶金鑰擴展機制隨機產生進階加密標準(Advanced Encryption Standard, AES)金鑰和毛蟲金鑰對(caterpillar key pairs)，並把進階加密標準金鑰和毛蟲公鑰放到 EeRaCertRequest，以及用註冊私鑰對 EeRaCertRequest 內容進行簽章，再發送 EeRaCertRequest 給登錄中心。登錄中心收到 EeRaCertRequest 後，驗證註冊憑證裡的註冊憑證中心簽章，並且以註冊憑證裡的註冊公鑰驗證 EeRaCertRequest 驗章；當驗章成功後，代表是合法的終端設備，並且根據終端設備的進階加密標準金鑰和毛蟲公鑰來產製繭公鑰，並且發送繭公鑰給授權憑證中心/假名憑證中心，由授權憑證中心/假名憑證中心根據繭公鑰產製蝴蝶公鑰和簽發授權憑證/假名憑證。其中，RaEeCertAck 是回應 EeRaCertRequest 的結果，並且由終端設備發起 EeRaDownloadRequest 下載授權憑證/假名憑證，最後由登錄中心把憑證資訊和授權憑證/假名憑證壓縮成 zip 檔傳送給終端設備(ITSC of IEEE VTS, 2022)。具體流程和實作效能比較可參考文獻(Chen et al., 2023)。

在憑證申請流程中，終端設備不與授權憑證中心/假名憑證中心直接連線，而是通過登錄中心，並且過程中由登錄中心擴展毛蟲公鑰為繭公鑰、由授權憑證中心/假名憑證中心擴展繭公鑰為蝴蝶公鑰，達到匿名性(ITSC of IEEE VTS, 2022)。

## 2.3 憑證格式

IEEE 1609.2 標準定義的憑證格式包含有版本(Version)、類型(Type)、簽發者(Issuer)、被簽署憑證內容(To Be Signed Certificate)、以及簽章(Signature)，如圖 3 所示(ITSC of IEEE VTS, 2023)。其中，雖然 IEEE 1609.2 標準定義的憑證類型包

含顯式憑證(explicit certificate)和隱式憑證(implicit certificate)(ITSC of IEEE VTS, 2023)，具體顯式憑證和隱式憑證的詳細比較可參考文獻(Chen, 2023)；由於已有研究指出後量子密碼學無法做出隱式憑證(Bindel & McCarthy, 2023)，所以本研究為公平比較，只考慮顯式憑證。

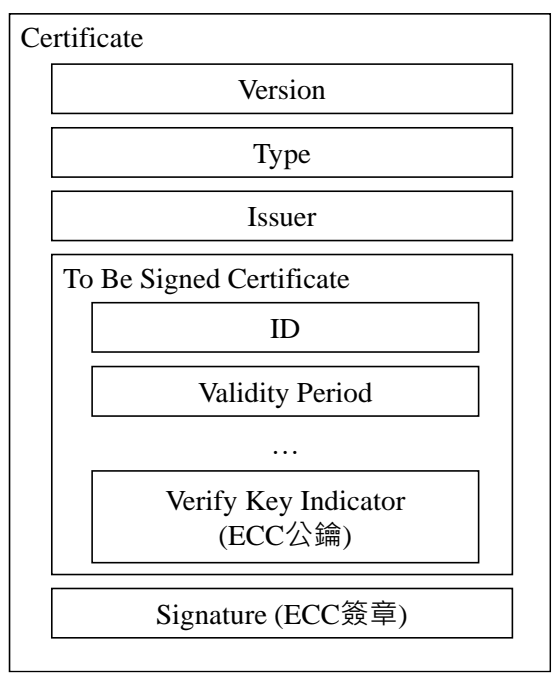

**圖 3　憑證格式(ITSC of IEEE VTS, 2023)**

　　以終端設備的註冊憑證為例，註冊憑證中的簽發者是註冊憑證中心憑證的雜湊值，用以表示由註冊憑證中心簽發，並且只存註冊憑證中心憑證雜湊值，減少憑證長度。被簽署憑證內容中驗證金鑰指示器(Verify Key Indicator, VKI)存放的是終端設備的註冊公鑰，在 IEEE 1609.2 標準中是 1 把橢圓曲線密碼學公鑰，即 1 個橢圓曲線點。最後，簽章欄位是存放由註冊憑證中心以註冊憑證中心橢圓曲線密碼學私鑰對被簽署憑證內容的簽章，即 1 個橢圓曲線點加上 1 個整數值。

　　以終端設備的授權憑證/假名憑證為例，授權憑證/假名憑證中的簽發者是授權憑證中心憑證/假名憑證中心憑證的雜湊值，用以表示由授權憑證中心/假名憑證中心簽發，並且只存授權憑證中心憑證/假名憑證中心憑證雜湊值，減少憑證長度。被簽署憑證內容中驗證金鑰指示器(Verify Key Indicator, VKI)存放的是終端設備的蝴蝶公鑰，在 IEEE 1609.2 標準中是 1 把橢圓曲線密碼學公鑰，即 1 個橢圓曲線點。最後，簽章欄位是存放由授權憑證中心/假名憑證中心以授權憑證中心/假名憑證中心橢圓曲線密碼學私鑰對被簽署憑證內容的簽章，即 1 個橢圓曲線點加上 1 個整數值。

　　由於實務上可壓縮橢圓曲線點以減少橢圓曲線密碼學公鑰長度和橢圓曲線

密碼學簽章長度,所以本研究後續做長度比較時皆採用已壓縮橢圓曲線點長度計算,未壓縮橢圓曲線點和已壓縮橢圓曲線點的差異可參考文獻(Chen, 2023)。

## 2.4 安全協定資料單元格式

IEEE 1609.2 標準定義的安全協定資料單元格式包含有被簽署資料(To Be Signed Data)、簽章者識別碼(Signer Identifier)、以及簽章(Signature),如圖 4 所示(ITSC of IEEE VTS, 2023)。

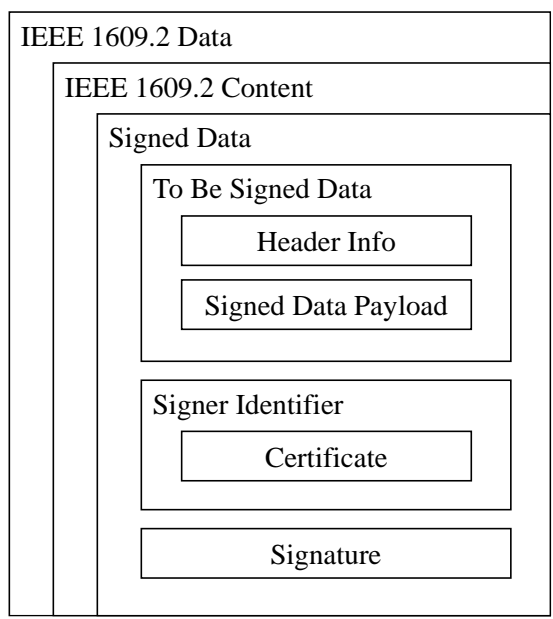

**圖 4 安全協定資料單元格式(ITSC of IEEE VTS, 2023)**

以終端設備的發送帶有基本安全訊息(Basic Safety Message, BSM)的安全協定資料單元為例,被簽署資料主要包含標頭資訊(Heander Info)和被簽署資料內容(Signed Data Payload),標頭資訊包含有提供者服務識別碼(Provider Service Identifier, PSID)、產製時間(Generation Time)等,而被簽署資料內容是智慧運輸系統(Intelligent Transportation System, ITS)訊息(即本例是基本安全訊息(Basic Safety Message, BSM))。簽章者識別碼主要包含簽章者(本例是終端設備)完整憑證或憑證摘要(digest),當帶完整憑證時,憑證長度將影響安全協定資料單元長度。最後,簽章欄位是存放由終端設備以蝴蝶私鑰(即 1 個橢圓曲線密碼學私鑰)對被簽署資料的簽章,即 1 個橢圓曲線點加上 1 個整數值(ITSC of IEEE VTS, 2023),具體實例可參考文獻(Chen et al., 2023)。

## 3. 後量子密碼學與橢圓曲線密碼學混合憑證方案

有鑑於現行的 IEEE 1609.2 標準(ITSC of IEEE VTS, 2023)和 IEEE 1609.2.1

後量子密碼學與橢圓曲線密碼學混合憑證方案—以車聯網安全憑證管理系統為例

標準(ITSC of IEEE VTS, 2022)主要採用橢圓曲線密碼學，所以將面臨量子計算攻擊的威脅。本研究提出後量子密碼學與橢圓曲線密碼學混合憑證方案，在 3.1 節介紹本研究提出的純後量子密碼學憑證和後量子密碼學與橢圓曲線密碼學混合憑證，並且進行比較與討論。在 3.2 節描述純後量子密碼學憑證適用於安全憑證管理系統的哪些角色，以及後量子密碼學與橢圓曲線密碼學混合憑證適用於安全憑證管理系統的哪些角色及情境。為保障終端設備的匿名性，本研究設計後量子密碼學與橢圓曲線密碼學混合憑證的匿名憑證流程，在 3.3 節介紹本研究提出的匿名憑證及其原理證明。

## 3.1 本研究提出的憑證格式與比較

本節提出純後量子密碼學憑證(如圖 5 所示)和後量子密碼學與橢圓曲線密碼學混合憑證(如圖 6 所示)，以下詳細說明兩種憑證格式。

純後量子密碼學憑證是指在驗證金鑰指示器改為存放憑證持有者的後量子密碼學公鑰，並且簽章也改為簽發者用其後量子密碼學私鑰對被簽署憑證內容的簽章，如圖 5 所示。以中繼憑證中心憑證為例，中繼憑證中心憑證裡的驗證金鑰指示器是中繼憑證中心的後量子密碼學公鑰，而簽章欄位是根憑證中心用根憑證中心的後量子密碼學私鑰對被簽署憑證內容的簽章。純後量子密碼學對抗量子計算攻擊的安全性較高，但由於憑證中需放置 1 個後量子密碼學公鑰和 1 個後量子密碼學簽章，將造成憑證長度過大並超過 1400 位元組，導致無法在車聯網環境中點對點(Peer-to-Peer)傳輸。

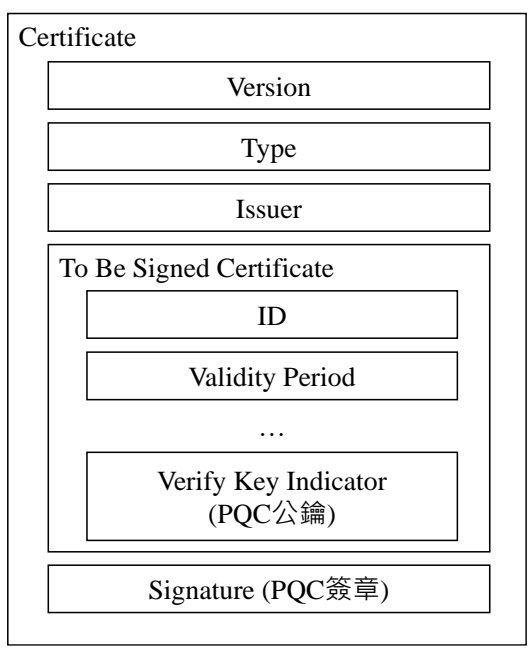

圖 5　純後量子密碼學憑證格式

　　為解決純後量子密碼學憑證長度過大和純橢圓曲線密碼學無法抗量子計算攻擊的問題，本研究設計後量子密碼學與橢圓曲線密碼學混合憑證。後量子密碼學與橢圓曲線密碼學混合憑證是指在驗證金鑰指示器仍維持存放憑證持有者的橢圓曲線密碼學公鑰，但簽章改為簽發者用其後量子密碼學私鑰對被簽署憑證內容的簽章，如圖 6 所示。以終端設備授權憑證為例，終端設備授權憑證裡的驗證金鑰指示器是終端設備的橢圓曲線密碼學公鑰，而簽章欄位是授權憑證中心/假名憑證中心用授權憑證中心/假名憑證中心的後量子密碼學私鑰對被簽署憑證內容的簽章。因此，後量子密碼學與橢圓曲線密碼學混合憑證中需放置 1 個橢圓曲線密碼學公鑰和 1 個後量子密碼學簽章，可以具有部分抗量子計算攻擊的能力，並且具有較短的封包長度，用以產製安全協定資料單元在車聯網環境中傳輸。

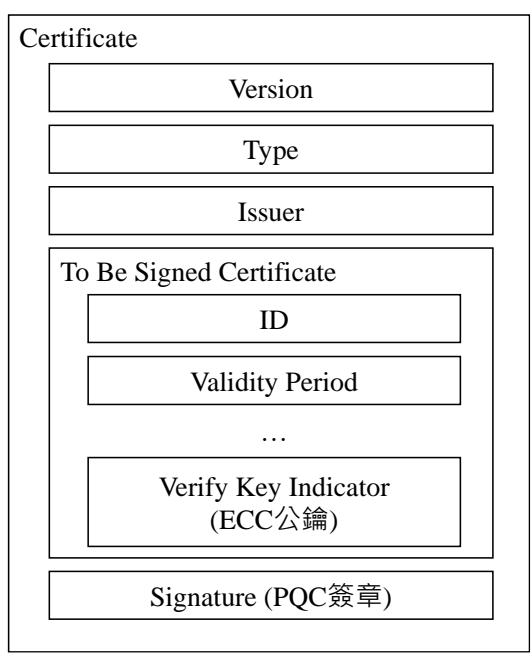

**圖 6　後量子密碼學與橢圓曲線密碼學混合憑證格式**

## 3.2 本研究安全憑證管理系統系統架構

　　本節將討論純後量子密碼學憑證和後量子密碼學與橢圓曲線密碼學混合憑證的適用性，對設計具備抗量子計算攻擊的車聯網安全憑證管理系統。在 3.2.1 節中說明本研究提出的憑證中心憑證方案；3.2.2 節和 3.2.3 節中分別討論本研究提出的終端設備註冊憑證方案和授權憑證/假名憑證方案。

### 3.2.1 憑證中心憑證和登錄中心憑證方案

　　安全憑證管理系統基礎設施中的伺服器主要包含有根憑證中心、中繼憑證中心、授權憑證中心/假名憑證中心、註冊憑證中心、以及登錄中心。由於這些伺

服器都可以離線簽發憑證,或是部分憑證中心和登錄中心可以通過 Internet 連線換發憑證,所以不存在封包長度限制。為提升抗量子計算攻擊的能力,純後量子密碼學憑證可以作為憑證中心憑證和登錄中心憑證方案,以防止安全憑證管理系統基礎設施被量子計算攻破。

### 3.2.2 終端設備註冊憑證方案

終端設備申請註冊憑證的流程主要是走 IEEE 1609.2.1 標準定義的憑證申請流程(如 2.2 節所述),是通過 Internet 連線到註冊憑證中心,而且在車聯網點對點傳輸的過程中不會用到註冊憑證。因此,終端設備註冊憑證可以不受限封包長度,所以可以把終端設備註冊憑證的抗量子計算攻擊能力提高,採用純後量子密碼學憑證可以作為終端設備註冊憑證方案。

### 3.2.3 終端設備授權憑證/假名憑證方案

終端設備申請授權憑證/假名憑證的流程主要是走 IEEE 1609.2.1 標準定義的憑證申請流程(如 2.2 節所述),雖然是通過 Internet 連線到授權憑證中心/假名憑證中心,但是在車聯網點對點傳輸的過程中將會用到授權憑證/假名憑證。因此,終端設備授權憑證/假名憑證需受到限封包長度的嚴格限制,不能超過 1400 位元組;然而,由於授權憑證中心憑證/假名憑證中心憑證已改為純後量子密碼學憑證,所以可以把終端設備授權憑證/假名憑證改為後量子密碼學與橢圓曲線密碼學混合憑證,以支援終端設備和授權憑證中心憑證/假名憑證中心之間通過後量子密碼學簽章和驗章達到抗量子計算攻擊的能力。

為科學化表示終端設備授權憑證/假名憑證長度進行比較,本研究定義公鑰度長度為 $k$ 個位元組、憑證中的簽章長度為 $s_1$ 個位元組、憑證中除了公鑰和簽章之外的資料長度(即包含版本、類型、簽發者、效期等)為 $c$ 個位元組、安全協定資料單元中的簽章長度為 $s_2$ 個位元組、以及安全協定資料單元中除了憑證和簽章之外的資料長度(即被簽署資料)為 $u$ 個位元組。因此,通過公式(1)可以計算終端設備授權憑證/假名憑證總長度為 $C$ 個位元組,並且通過公式(2)可以計算安全協定資料單元總長度為 $U$ 個位元組。

$$C = c + k + s_1 \tag{1}$$

$$U = u + C + s_2 \tag{2}$$

雖然在 IEEE 1609.2 標準定義的安全協定資料單元可只帶憑證摘要,而不帶

完整憑證(ITSC of IEEE VTS, 2023);然而,在實務上仍需要發送帶有完整憑證的安全協定資料單元,在北美車聯網概念性驗證場域中,要求車載設備每 450 毫秒發送一筆發送帶有完整憑證的安全協定資料單元(V2X Core Technical Committee, SAE, 2020)。具體各個密碼學演算法的終端設備授權憑證/假名憑證和安全協定資料單元長度比較將在 4.2 節中討論。

## 3.3 匿名憑證流程設計和原理證明

由於研究指出後量子密碼學無法做出隱式憑證(Bindel & McCarthy, 2023),但在車聯網車輛隱私是必要克服的挑戰。因此,本研究在 IEEE 1609.2.1 標準的蝴蝶金鑰擴展機制(ITSC of IEEE VTS, 2022)基礎上提出基於後量子密碼學與橢圓曲線密碼學混合憑證的匿名憑證,為 3.2.3 節中提出的終端設備授權憑證/假名憑證方案提供匿名性。在 3.3.1 節中將介紹本研究提出的基於後量子密碼學與橢圓曲線密碼學混合憑證的匿名憑證流程,並且在 3.3.2 節從原理上證明其可行性。

### 3.3.1 匿名憑證流程設計

本研究提出的基於後量子密碼學與橢圓曲線密碼學混合憑證的匿名憑證流程,由終端設備產製毛蟲金鑰對,再由登錄中心依毛蟲公鑰擴展為繭公鑰,最後再由授權憑證中心憑證/假名憑證中心依繭公鑰擴展為蝴蝶公鑰,如圖 7 所示。

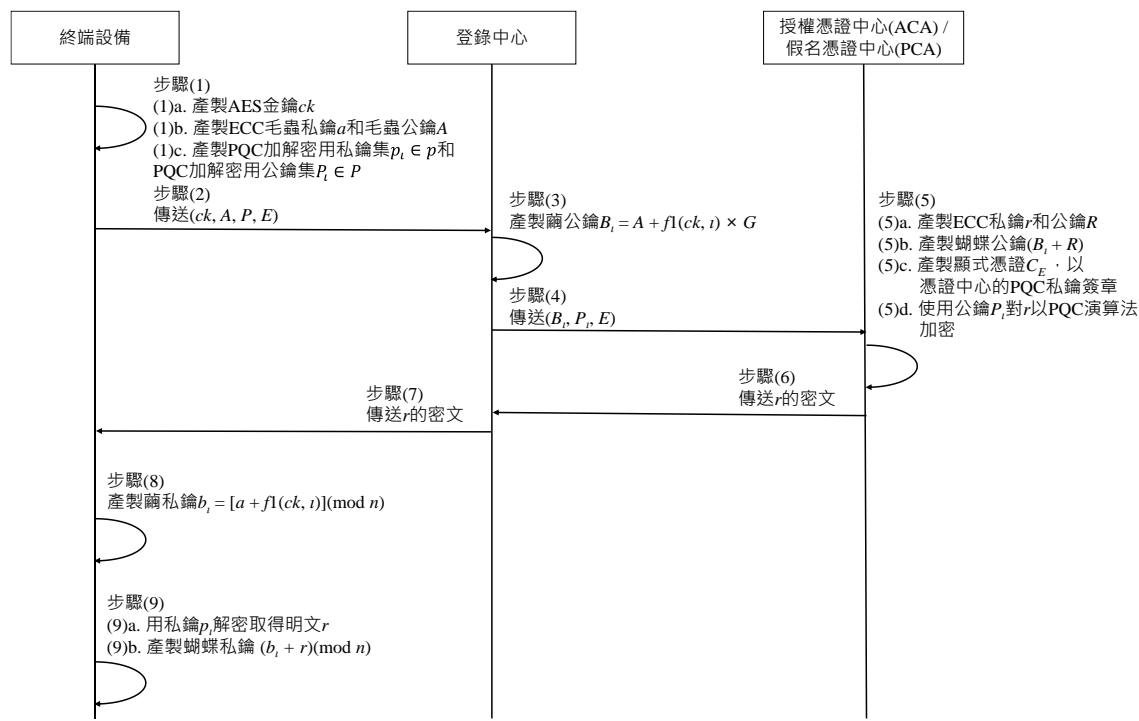

**圖 7　基於後量子密碼學與橢圓曲線密碼學混合憑證的匿名憑證**

本研究提出基於後量子密碼學與橢圓曲線密碼學混合憑證的匿名憑證流程說明如下：

(1). 由終端設備產製 AES 金鑰、ECC 金鑰對、PQC 金鑰對，具體方法流程包含有：

    a. 產製 AES 金鑰 $ck$，作為簽章使用。其中，參數 $ck$ 是對稱式金鑰。

    b. 產製 ECC 金鑰對 $A = aG$，作為毛蟲金鑰對，簽章使用。其中，參數 $a$ 是私鑰、參數 $A$ 是公鑰、參數 $G$ 是橢圓曲線的基準點。

    c. 產製 PQC 加解密用金鑰對集$(p, P)$，$p$ 作為加解密用私鑰集，$P$ 作為加解密用公鑰集。其中，第 $\iota$ 對金鑰對表示為$(p_\iota, P_\iota)$，$p_\iota \in p$、$P_\iota \in P$。

(2). 由終端設備將產製的對稱式金鑰 $ck$、毛蟲公鑰 $A$、加解密用公鑰集 $P$、以及被簽署憑證內容 $E$ 發送給登錄中心。

(3). 由登錄中心基於毛蟲公鑰產製繭公鑰 $B_\iota = A + f1(ck, \iota) * G$。其中，參數 $\iota$ 是增量整數；函數 $f1$ 是基於 AES 加密演算法的擴展函數，可運用 AES 金鑰 $ck$ 加密參數 $\iota$ 值，得到整數密文，避免被攻擊者取得參數 $\iota$ 明文的情況下從繭公鑰推導出毛蟲公鑰。

(4). 由登錄中心將產製的繭公鑰 $B_\iota$、毛蟲公鑰 $A$、加解密用公鑰 $P_\iota$、以及被簽署憑證內容 $E$ 發送給授權憑證中心憑證/假名憑證中心。

(5). 授權憑證中心憑證/假名憑證中心基於繭公鑰產製蝴蝶公鑰，蝴蝶公鑰可以作為匿名憑證的公鑰驗章使用，具體方法流程包含有：

    a. 產製 ECC 金鑰對 $R = rG$。其中，參數 $r$ 是私鑰、參數 $R$ 是公鑰、參數 $G$ 是橢圓曲線的基準點。

    b. 產製蝴蝶公鑰$(B_\iota + R)$。

    c. 產製終端設備授權憑證/假名憑證 $C_E$，以授權憑證中心憑證/假名憑證中心的 PQC 私鑰簽章。

    d. 運用加解密用公鑰 $P_\iota$ 以後量子密碼學演算法對 $r$ 加密和簽章，其中 $r$ 的密文係 $r'$。

(6). 授權憑證中心憑證/假名憑證中心發送密文 $r'$ 和簽章給登錄中心。

(7). 登錄中心密文 $r'$ 給終端設備。

(8). 終端設備根據 $\iota$ 值產製繭私鑰 $b_\iota = [a + f1(ck, \iota)](\text{mod } n)$。其中，函數 $f1$ 是基於 AES 加密演算法的擴展函數，可運用 AES 金鑰 $ck$ 加密參數 $\iota$ 值，得到整數密文、參數 $n$ 是橢圓曲線的階。

(9). 終端設備使用加解密用私鑰 $p_\iota$ 解密取得 $r$ 值，並且運用繭私鑰 $b_\iota$ 和 $r$ 值產製蝴蝶私鑰，蝴蝶私鑰可作為匿名憑證的私鑰簽章使用，具體方法流程包含有：

   a. 運用加解密用私鑰 $p_\iota$ 解密密文 $r'$，取得明文 $r$。

   b. 產製蝴蝶私鑰 $(b_\iota + r)(\text{mod } n)$。其中，參數 $n$ 是橢圓曲線的階。

### 3.3.2 匿名憑證可行性分析和匿名性分析

本節將分別討論匿名憑證可行性和匿名憑證匿名性。

**匿名憑證可行性分析**：本研究設計的基於後量子密碼學與橢圓曲線密碼學混合憑證的匿名憑證流程可以建構在 IEEE 1609.2.1 標準的憑證申請流程(如 2.2 節所述)(ITSC of IEEE VTS, 2022)上執行。其中，雖然本研究提出基於後量子密碼學與橢圓曲線密碼學混合憑證的匿名憑證流程需要包含後量子密碼學公鑰集(如圖 7 中的步驟(1).c)，所以在發出 EeRaCertRequest 時的資料長度較大；然而，由於終端設備和登錄中心之間的通訊主要通過 Internet 連線，而不是車聯網點對點傳輸，所以不受封包長度限制。因此，憑證申請流程可以正常執行，並且最後在 EeRaDownloadRequest 後取得由授權憑證中心憑證/假名憑證中心簽發帶有蝴蝶公鑰的終端設備授權憑證/假名憑證作為匿名憑證。

**匿名憑證匿名性分析**：在本研究設計的基於後量子密碼學與橢圓曲線密碼學混合憑證的匿名憑證流程中，因為登錄中心不知道明文 $r$ 值和 $R$ 值，所以無法從蝴蝶公鑰推導對應的繭公鑰；此外，登錄中心亦不知毛蟲私鑰，所以也無法推導繭私鑰和蝴蝶私鑰。對授權憑證中心憑證/假名憑證中心而言，因為授權憑證中心憑證/假名憑證中心不知道擴展函數加密後的 $\iota$ 值，所以無法從繭公鑰推導對應的毛蟲公鑰。此外，授權憑證中心亦不知毛蟲私鑰，所以也無法推導繭私鑰和蝴蝶私鑰。因此，可以保障車聯網中的其他設備(包含登錄中心和憑證中心)無法得知毛蟲公鑰和蝴蝶公鑰的對應關係，從而證實匿名性。

## 4. 實驗比較與討論

本節將實作和驗證本研究提出的後量子密碼學與橢圓曲線密碼學混合憑證方案和匿名憑證流程，並且驗證和比較其效能。4.1 節介紹本研究的實驗環境和

實驗場域；4.2 節、4.3 節、以及 4.4 節分別從憑證長度與安全協定資料單元長度、簽章和驗章效率、加密和解密效率等不同面向來比較本研究方法的效能。最後，4.5 節總結實驗結果和提供未來實施車聯網安全憑證管理系統的建議。

## 4.1 實驗環境與實驗場域

在實驗環境中，本研究採用中華電信的安全憑證管理系統和公信電子(Clientron)的車載設備在新北淡海新市鎮場域進行實證。其中，在 OmniAir 748 文件中指出中華電信的安全憑證管理系統符合 IEEE 1609.2 和 IEEE 1609.2.1 標準及其規範(Conley, 2023)。除此之外，中華電信安全憑證管理系統和公信電子車載設備已在新北淡海新市鎮場域於 2023 年 11 月「2023 OmniAir Taipei PlugFest 臺北聯網車互通測試大會」進行實際互測通過(OmniAir, 2023)。本研究在上述基礎中加入後量子密碼學與橢圓曲線密碼學混合憑證方案，並採集真實資料進行驗證。其中，本研究採用的公信電子車載設備的作業系統是 Android 10，處理器是聯發科 Autus I20 (MT2712)，程式採用 Java 語言開發。

現行的 IEEE 1609.2 標準(ITSC of IEEE VTS, 2023)和 IEEE 1609.2.1 標準(ITSC of IEEE VTS, 2022)採用的簽驗章演算法是橢圓曲線數位簽章算法(Elliptic Curve Digital Signature Algorithm, ECDSA)，並且橢圓曲線參數採用美國國家標準與技術研究院定義的 P-256，所以本研究把 ECDSA P-256 作為比較基準。在後量子密碼學的部分，目前美國國家標準與技術研究院已經評選出 Dilithium、Falcon、以及 SPHINCS+ 三套後量子密碼學簽驗章演算法標準；為最小化憑證長度，本研究主要對這三套後量子密碼學簽驗章演算法的最小參數組合，包含 Dilithium-2、Falcon-512、以及 SPHINCS+ SHA2-128f (Alagic et al., 2022)。

在安全等級的部分，採用美國國家標準與技術研究院用以衡量後量子密碼學演算法的定義(NIST, 2023)。因此橢圓曲線密碼學屬於不具備抗量子計算攻擊的能力，所以不屬於任何安全等級級別。

## 4.2 憑證長度與安全協定資料單元長度比較

本節根據 3.2.3 節和 4.1 節中定義的參數進行實驗比較，結果如表 1 所示。其中，根據統計公信電子車載設備發送基本安全訊息時，憑證中除了公鑰和簽章之外的資料長度(即包含版本、類型、簽發者、效期等) $c$ 為 34 個位元組、安全協定資料單元中除了憑證和簽章之外的資料長度(即被簽署資料) $u$ 為 68 個位元組，視為常數值來加總和比較。而公鑰度長度 $k$、憑證中的簽章長度 $s_1$、安全協定資料單元中的簽章長度 $s_2$ 將根據不同的密碼學演算法而異，再根據公式(1)計

算終端設備授權憑證/假名憑證總長度 $C$ 和公式(2)計算安全協定資料單元總長度 $U$。

由實驗結果可知，現行的 IEEE 1609.2 標準(ITSC of IEEE VTS, 2023)和 IEEE 1609.2.1 標準(ITSC of IEEE VTS, 2022)採用的簽驗章演算法 ECDSA P-256 具有最較短的公鑰長度(即 33 位元組)和簽章長度(即 65 位元組)，所以具有最短的終端設備授權憑證/假名憑證長度和最短的安全協定資料單元長度，但不具備抗量子計算攻擊的能力，所以將被量子電腦輕易破解。

表 1　憑證長度與安全協定資料單元長度比較(單位：位元組)

| 安全等級 | 模式 | 憑證中心演算法 | 終端設備演算法 | $k$ | $s_1$ | $c$ | $C$ | $s_2$ | $u$ | $U$ |
|---|---|---|---|---|---|---|---|---|---|---|
| 不安全 | 純 ECC (IEEE 標準) | ECDSA P-256 | ECDSA P-256 | 33 | 65 | 34 | 132 | 65 | 68 | 265 |
| 2 | 純 PQC | Dilithium-2 | Dilithium-2 | 1312 | 2420 | 34 | 3766 | 2420 | 68 | 6254 |
| 1 | 純 PQC | Falcon-512 | Falcon-512 | 898 | 666 | 34 | 1598 | 666 | 68 | 2332 |
| 1 | 純 PQC | SPHINCS+ SHA2-128f | SPHINCS+ SHA2-128f | 33 | 16,720 | 34 | 16787 | 16,720 | 68 | 33575 |
| 2 | PQC+ECC 混合 | Dilithium-2 | ECDSA P-256 | 33 | 2420 | 34 | 2487 | 65 | 68 | 2620 |
| 1 | PQC+ECC 混合 | Falcon-512 | ECDSA P-256 | 33 | 666 | 34 | 733 | 65 | 68 | 866 |
| 1 | PQC+ECC 混合 | SPHINCS+ SHA2-128f | ECDSA P-256 | 33 | 16,720 | 34 | 16787 | 65 | 68 | 16920 |

從實驗結果可以觀察到由於目前通過美國國家標準與技術研究院評選為標準的後量子密碼學簽驗章演算法，其公鑰長度和簽章長度都較大，所以純後量子密碼學憑證不適用於車聯網點對點傳輸。例如：後量子簽驗章演算法 Dilithium-2 的公鑰長度 1312 位元組和簽章長度 2420 位元組，所以終端設備授權憑證/假名憑證長度和安全協定資料單元長度分別為 3766 位元組和 6254 位元組，超過封包長度 1400 位元組的限制。因此，純後量子密碼學憑證僅能適用於 Internet 傳輸

後量子密碼學與橢圓曲線密碼學混合憑證方案─以車聯網安全憑證管理系統為例

時,所以適用於各個憑證中心的憑證、登錄中心的憑證、以及終端設備的註冊憑證。

在後量子密碼學與橢圓曲線密碼學混合憑證方案中,可以觀察由於後量子簽驗章演算法 Falcon-512 的簽章長度(即 666 位元組)較其他後量子簽驗章演算法的簽章短。因此,終端設備授權憑證/假名憑證存放憑證中心的 Falcon-512 簽章(即長度 666 位元組)和終端設備的 ECDSA P-256 公鑰(即長度 33 位元組),憑證總長度只要 733 位元組。並且,由於後續終端設備簽發安全協定資料單元時,是存放 ECDSA P-256 簽章 (即長度 65 位元組),所以安全協定資料單元長度為 866 位元組,可以符合封包長度 1400 位元組的限制。

為全面比較各種後量子密碼學演算法參數及其對應的長度,表 2 以美國國家標準與技術研究院評選出來的後量子密碼學演算法進行比較。由實驗結果可知,由於 ECDSA P-256 擁有最短的公鑰長度和簽章長度,所以長度可以符合車聯網點對點傳輸的需求;然而,ECDSA P-256 是基於橢圓曲線密碼學,所以將可能被量子計算攻擊破解,所以不具安全性。SPHINCS+演算法雖然有較短的公鑰長度,然而簽章長度太大,超過 1400 位元組,所以不適用於車聯網。Dilithium-2 的公鑰長度雖能小於過 1400 位元組,但安全協定資料單位還需包含其他資訊,所以將超出封包長度限制。因此,僅有 Falcon-512 能適用於現行的車聯網環境。

表 2　各個密碼學演算法公鑰長度和簽章長度比較(單位:位元組)

| 演算法 | 安全等級 | 公鑰長度(byte) | 簽章長度(byte) |
| --- | --- | --- | --- |
| ECDSA P-256 | 不安全 | 33 | 64 |
| Dilithium-2 | 2 | 1,312 | 2,420 |
| Dilithium-3 | 3 | 1,952 | 3,309 |
| Dilithium-5 | 5 | 2,592 | 4,627 |
| Falcon-512 | 1 | 897 | 666 |
| Falcon-1024 | 5 | 1793 | 1280 |
| SPHINCS+ SHA2-128f | 1 | 32 | 16,720 |
| SPHNICS+ SHA2-192f | 3 | 48 | 34,896 |
| SPHINCS+ SHA2-256f | 5 | 64 | 49,312 |

## 4.3 簽章和驗章效率比較

本節將比較全部已被美國國家標準與技術研究院評選為標準的後量子密碼學數位簽章演算法，包含 Dilithium、Falcon、以及 SPHINCS+ (Alagic et al., 2022)。以及比較部分參與評選但未入選為標準的後量子密碼學數位簽章演算法，包含 GeMSS、Picnic、以及 Rainbow (Alagic et al., 2022)。本研究將上述演算法實作到公信電子車載設備上，進行實證分析和比較；然而，由於演算法無法支援各種安全等級(例如：Dilithium 只支持安全等級 2、3、5，Falcon 只支援安全等級 1、5)，所以只挑最小安全等級進行比較，以盡可能符合車聯網點對點傳輸的需求。

各種後量子密碼學數位簽章演算法的產製金鑰時間、產製簽章時間、驗證簽章時間結果如表 3 所示。由於產製金鑰可以批次處理或離峰時段進行，所以產製金鑰時間在車聯網應用中較沒有急迫性。然而，在北美車聯網概念性驗證場域(新北淡海新市鎮場域也採用相同限制)中，要求車載設備每 100 毫秒發送一筆帶有基本安全訊息的安全協定資料單元，所以產製簽章時間和驗證簽章時間應該在 100 毫秒內完成。由實驗結果可知，雖然 SPHINCS+ SHA2-128f 和 GeMSS BlueGeMSS128 的驗證簽章時間可以符合需求，但產製簽章時間太長，所以不適用於車聯網環境。此外， Picnic-1fs 和 Rainbow-III-Classic 的產製簽章時間和驗證簽章時間皆超過 100 毫秒，所以也無法應用於車聯網概念性驗證場域。因此，只有 Dilithium-2、Falcon-512 可適用於車聯網，並且 Falcon-512 的驗證簽章時間最短僅需 0.87 毫秒，更適合在需要驗證大量車聯網封包的情境。在同時考慮憑證長度和安全協定資料單位長度限制下，Falcon-512 是最合適的方案。

表 3　簽章和驗章效率比較(單位：毫秒)

| 簽驗章演算法 | 安全等級 | 產製金鑰時間 | 產製簽章時間 | 驗證簽章時間 |
| --- | --- | --- | --- | --- |
| ECDSA P-256 | 不安全 | 18.38 | 18.20 | 26.51 |
| Dilithium-2 | 2 | 8.72 | 12.27 | 4.20 |
| Falcon-512 | 1 | 177.30 | 46.88 | 0.87 |
| SPHINCS+ SHA2-128f | 1 | 43.83 | 915.95 | 55.75 |
| GeMSS BlueGeMSS128 | 1 | 11039.04 | 176516.12 | 6.33 |
| Picnic-1fs | 1 | 8.72 | 183.46 | 124.42 |
| Rainbow-III-Classic | 3 | 4396.03 | 435.92 | 125.86 |

## 4.4 加密和解密效率比較

　　本節將比較已被美國國家標準與技術研究院評選為標準的後量子密碼學金鑰封裝演算法(即加密和解密使用)，僅有 Kyber (Alagic et al., 2022)。並且比較進入第 4 輪評選尚未成為標準的後量子密碼學金鑰封裝演算法，包含 BIKE、Classic McEliece、HQC (Alagic et al., 2022)。以及比較部分參與評選但未入選為標準的後量子密碼學金鑰封裝演算法，包含 Frodo、NTRU、以及 SABER (Alagic et al., 2022)。本研究將上述演算法實作到公信電子車載設備上，只挑最小安全等級進行比較，以盡可能符合車聯網點對點傳輸的需求。

　　各種後量子密碼學金鑰封裝演算法的產製金鑰時間、加密時間、解密時間結果如表 4 所示。如 4.3 節所述，產製金鑰時間在車聯網應用中較沒有急迫性，所以比較沒有嚴格限制時間。然而，在北美車聯網概念性驗證場域(新北淡海新市鎮場域也採用相同限制)中，要求車載設備每 100 毫秒發送一筆帶有基本安全訊息的安全協定資料單元，如果內容需要加密和解密，則加密時間和解密時間應該在 100 毫秒內完成。由實驗結果可知，雖然後量子密碼學數位簽章演算法 McEliece-348864 的加密時間最短僅 0.99 毫秒，但解密時間較長達 29.58 毫秒。其中，後量子密碼學金鑰封裝演算法 Kyber-512、NTRU-hps2048509、SABER-348864 皆為基於格(lattice-based)密碼學，具有較短的加密時間和解密時間，皆為 10 毫秒內。因此，同屬於基於格密碼學系列，所以美國國家標準與技術研究院綜合評估後選定 Kyber 為標準。因此，未來在車聯網環境的後量子密碼學金鑰封裝演算法可採用 Kyber 演算法。

表 4　加密和解密效率比較(單位：毫秒)

| 加解密演算法 | 安全等級 | 產製金鑰時間 | 加密時間 | 解密時間 |
|---|---|---|---|---|
| Kyber-512 | 1 | 4.84 | 4.17 | **<u>3.96</u>** |
| BIKE-128 | 1 | 40.93 | 3.33 | 29.26 |
| McEliece-348864 | 1 | 4845.38 | **<u>0.99</u>** | 29.58 |
| HQC-128 | 1 | 13.99 | 27.85 | 42.06 |
| Frodo-KEM-640 | 1 | 65.95 | 72.75 | 66.68 |
| NTRU-hps2048509 | 1 | 70.12 | 2.64 | 6.70 |
| SABER-348864 | 1 | 5.18 | 4.15 | 4.74 |

# 5.結論與建議

有鑑於車聯網挑戰包含封包長度、簽章和驗章效率、安全等級、以及車輛隱私，本研究提出的後量子密碼學與橢圓曲線密碼學混合憑證方案，搭配本研究提出的匿名憑證流程可以滿足前述的車聯網挑戰。

(1). **封包長度**：本研究提出的後量子密碼學與橢圓曲線密碼學混合憑證方案，在憑證中心採用後量子密碼學演算法 Falcon-512 且終端設備採用橢圓曲線密碼學演算法 ECDSA P-256 搭配的情況下，安全協定資料單位長度僅需 866 位元組，低於 1400 位元組隊制。

(2). **簽章和驗章效率**：本研究實作各種後量子密碼學演算法在公信電子車載設備於新北淡海新市鎮場域實證。實驗結果顯示後量子密碼學演算法 Falcon-512 驗證簽章時間僅需 0.87 毫秒比現行標準的 ECDSA P-256 驗證簽章時間(即 26.51 毫秒)來得短。

(3). **安全等級**：由於現行標準採用橢圓曲線密碼學，已經不安全，所以本研究提出後量子密碼學與橢圓曲線密碼學混合憑證方案可以滿足美國國家標準與技術研究院定義的安全等級。其中，後量子密碼學數位簽章演算法可以採用 Falcon-512，而後量子密碼學金鑰封裝演算法可以採用 Kyber-512。

(4). **車輛隱私**：為保護車輛隱私，本研究提出匿名憑證流程，可以讓其他設備(包含憑證中心和登錄中心)都無法推導蝴蝶公鑰和毛蟲公鑰的關聯，從而提供匿名性。

由於本研究提出的純後量子密碼學憑證方案保障了車聯網安全憑證管理系統在憑證中心憑證、登錄中心憑證、以及終端設備註冊憑證，而本研究提出的後量子密碼學與橢圓曲線密碼學混合憑證方案雖然保障了終端設備授權憑證/假名憑證，但在安全協定資料單元仍是採用橢圓曲線密碼學簽章，所以未來可以通過高頻率更新授權憑證/假名憑證來避免被攻擊。

除此之外，美國國家標準與技術研究院目前仍在徵選簽章長度短且簽章快速的後量子密碼學數位簽章演算法。未來如果評選出簽章長度更短且簽章更快速的演算法，可以採用新的標準後量子密碼學數位簽章演算法在終端設備授權憑證/假名憑證建立純後量子密碼學憑證方案，從而保護安全協定資料單元的簽章，讓整個車聯網安全憑證管理系統都建構在後量子密碼學的保障下來抵抗量子計算攻擊。

後量子密碼學與橢圓曲線密碼學混合憑證方案—以車聯網安全憑證管理系統為例# 參考文獻